\documentclass[twocolumn]{aastex631}

\usepackage{amsmath}
\usepackage{float}
\usepackage{multirow}
\usepackage{textcomp}
\usepackage{mathrsfs}
\usepackage{rsfso}
\usepackage[caption=false]{subfig}

\newcommand{\instrname}{MOMOS}
\newcommand{\instrnamecaps}{MOMOS}
\DeclareUnicodeCharacter{2061}{}

\begin{document}
\captionsetup[subfigure]{labelformat=empty}

\title{\instrname: The Multi-Object MKID Optical Spectrometer Simulator and Data Reduction Package}

\correspondingauthor{Crystal S. Kim}
\email{crystalkim@ucsb.edu}

\author[0000-0002-5683-5950]{Crystal S. Kim}
\affiliation{Department of Physics \\
University of California Santa Barbara \\
Santa Barbara, CA 93106, USA}

\author[0000-0002-4272-263X]{John I. Bailey, III}
\affiliation{Caltech Optical Observatories \\
California Institute of Technology \\
Pasadena, CA 91125, USA}

\author[0000-0002-2019-4995]{Ronald A. L\'opez}
\affiliation{Department of Physics \\
University of California Santa Barbara \\
Santa Barbara, CA 93106, USA}

\author[0000-0003-3568-7634]{W. Hawkins Clay}
\affiliation{Department of Physics \\
University of California Santa Barbara \\
Santa Barbara, CA 93106, USA}

\author[0000-0003-0526-1114]{Benjamin A. Mazin}
\affiliation{Department of Physics \\
University of California Santa Barbara \\
Santa Barbara, CA 93106, USA}

\begin{abstract}
\instrname, the Multi-Object MKID Optical Spectrometer, is a proposed visible wavelength spectrometer that uses MKIDs (Microwave Kinetic Inductance Detectors) targeting an initial resolving power of 3500. With their modest wavelength-resolving abilities, MKIDs take the place of both the cross disperser and detector in the spectrometer. MKIDs lack read noise and dark current enabling noiseless post-observation rebinning and characterization of faint objects, as well as time-resolved photon-counting spectroscopy. This work presents an \instrname\ simulator customizable for different \instrname\ configurations. Treating simulator products as inputs, an algorithm was developed and implemented in the \instrname\ data reduction package to calibrate and extract spectra.
\end{abstract}


\section{Introduction} \label{sec:intro}
Most astronomical spectrometers use an echelle grating followed by a cross-disperser and detector to analyze luminous sources for emission and absorption features, usually across multiple spectral orders to yield a wide wavelength coverage. They can be found at almost every major telescope facility \citep{rayner_spex_1998, baudrand_fiber_2000, dressler_science_2003, tokoku_moircs_2003, crane_fan_2005, hill_virus_2006, edelstein_tedi_2007, flaugher_dark_2014, crause_spupnic_2016} and are a mainstay instrument in astronomy. An advantage of the echelle spectrometer is the ability to use rectangular detectors already widely used in other astrophysical applications, most commonly charge-coupled devices (CCDs). CCDs can scale to Gigapixel arrays with small, energy-efficient, and sensitive pixels.

Proposals to use energy-resolving detectors in echelle spectrometers date back over 20 years~\citep{cropper_concept_2003}. The spectrometer described in this paper is based off of a similar conceptual instrument called KIDSpec (Kinetic Inductance Detector Spectrograph) from \citet{obrien_kidspec_2014} that uses an emerging detector technology known as Microwave Kinetic Inductance Detectors (MKIDs). These photon counting, superconducting detectors resolve both the energy and arrival time for each photon \citep{day_broadband_2003}, lack both read noise and dark current, and are largely unaffected by cosmic ray contamination after post-processing.

Several low-resolution MKID integral field spectrographs (IFS) exist today \citep{mazin_arcons_2013, meeker_darkness_2018, walter_mkid_2020, swimmer_xkid_2023}. These IFS's return spectral information solely using the innate wavelength discrimination of MKIDs. 
With $\mathcal{R}_{\textrm{MKID}}\equiv\lambda/d\lambda\lesssim 12$ at 600~nm, these instruments largely behave as broadband integrated photometers with high temporal resolution. Although UVOIR MKID device fabrication remains a significant challenge, $\mathcal{R}_{\textrm{MKID}}\approx46$ at 600~nm has been measured \citep{de_visser_phonon-trapping-enhanced_2021} with a maximum theoretical $\mathcal{R}_{\textrm{MKID}}\sim$140 at 600~nm for PtSi \citep{zobrist_membraneless_2022}. All further $\mathcal{R}_{\textrm{MKID}}$ are referenced with respect to 600~nm.

These higher $\mathcal{R}_{\textrm{MKID}}$ values more than suffice to discriminate spectral orders when used in moderate to high-resolution ($\mathcal{R}_{\textrm{spec}}\sim 4000$ to $100,000$) echelle(tte) spectrometers and would eliminate the need for a cross-dispersing optical element. With $\mathcal{R}_{\textrm{MKID}}=15$, an MKID can discriminate orders 5 through 9 in the 400 to 800~nm for an $\mathcal{R}_{\textrm{spec}}\sim3500$. Because light is dispersed in only one direction (no cross-disperser), only a single row of MKIDs is required for the read out of a spectrum. By strategically locating sources and employing multiple linear MKID pixel arrays, a highly pixel-efficient, compact, multi-object spectrometer capable of simultaneously obtaining all spectral orders becomes possible.

This work introduces a simulator and data reduction package for \instrname, an MKID spectrometer testbed for the exploration of multi-object echelle designs employing MKID detectors. Such instruments offer significant potential to astronomy as they capture the entire echellogram for every fiber without either the read noise or dark current of traditional CCD or CMOS arrays. This fundamentally alters the design-space by enabling digital re-binning to match resolving power to source brightness and science case while simultaneously eliminating the wavelength coverage/target-count trade of existing multi-object spectrometers.
Moreover, the temporal resolution of MKIDs can enable fundamentally new spectroscopic analysis approaches for stellar astrophysics. 

As a testbed, \instrname\ is expected to undergo grating and resolution upgrades. Its initial incarnation is as a medium-resolution MKID spectrometer operating from 400-800 nm with up to five fiber feeds, an echellette grating in Littrow configuration, an MKID device with five 2048-pixel linear MKID arrays with 20x200 micron pixels, and ability to discriminate orders 4-7 in a commercially-available off-blaze grating configuration. Further changes will further expand \instrname's scientific value and inform the design of future MKID-based spectrometers. The \instrname\ simulator has been instrumental in determining how future \instrname\ upgrades might impact data reduction capabilities. It also produced the realistic input data used to test the data reduction package. Likewise, real \instrname\ output will be processed through the data reduction described in this paper.

A simulation of KIDSpec, the instrument concept which \instrname\ was based on, was recently developed to understand improvements KIDSpec could bring to low-SNR spectroscopy \citep{hofmann_ksim_2023}. There are several key differences between this work and the KIDSpec simulation, known as KSIM. The \instrname\ simulator needs to mimic realistic MKID data as much as possible. This means that all wavelengths are converted to an appropriate phase response and then multiplied with random offsets on a pixel to pixel basis, which makes wavelength or order information impossible to recover without further processing. Conversely, KSIM retains order and wavelength information throughout the simulation and reduction. Secondly, since real instruments are not always perfectly aligned like their optical models, it cannot be assumed which portion of the spectrum (i.e., which photon energies) are incident on which pixels. That is why the \instrname\ simulator involves a wavelength calibration step that is not present in KSIM. Secondly, one of the concerns mentioned by \citet{hofmann_ksim_2023} was the degree of order misidentification due to the overlap of order Gaussians, which limited KSIM to $\mathcal{R}_{\textrm{MKID}}\approx 22$. Because of the order-bleeding (overlap) correction described here in Section~\ref{subsec:bleed}, the \instrname\ data reduction is suitable down to $\mathcal{R}_{\textrm{MKID}}=15$. These major differences highlight the intended purpose of either simulator; KSIM characterizes achievable science goals as a high fidelity SNR calculator over a range of conditions and the \instrname\ simulator provides realistic data to rigorously test the data reduction before it is used with real \instrname\ data.

The \instrname\ simulator takes an input spectrum and yields photon events as time-tagged MKID resonator phase shifts, which the data reduction package processes into tabulated photon data and standard astrophysical spectral orders. The simulator is described in Section~\ref{sec:sim} and the \instrname\ data reduction in Section~\ref{sec:extract}.

\section{\instrnamecaps\ Simulator} \label{sec:sim}
The \instrname\ simulator takes a model input spectrum and applies the following effects: telluric attenuation, addition of sky emission lines, multiplication with grating blaze function, convolution with the optical Line Spread Function, convolution with the MKID response function, conversion to phase response, and storage to MKID photon table object. A full simulator schematic of steps and options is shown in Appendix~\ref{sec:simoptions}.

\subsection{Model Spectra} \label{subsec:models}
The user initiates the \instrname\ simulator by indicating one of the included spectra options (PHOENIX model \citep{husser_new_2013}, blackbody, emission lamp, flat-field, and SkyCalc telluric emission \citep{noll_atmospheric_2012}) or by supplying an input spectrum. In addition, the user may select the option to alter the input spectrum with added telluric emission and multiplication of telluric throughput to simulate a ground-based observation. The spectrum, denoted as $S(\lambda)$, is further attenuated by instrument-specific filters.

\subsection{Blaze Function} \label{subsec:grating}
\begin{figure*}
\centering
    \subfloat{
    \includegraphics[width=0.5\linewidth]{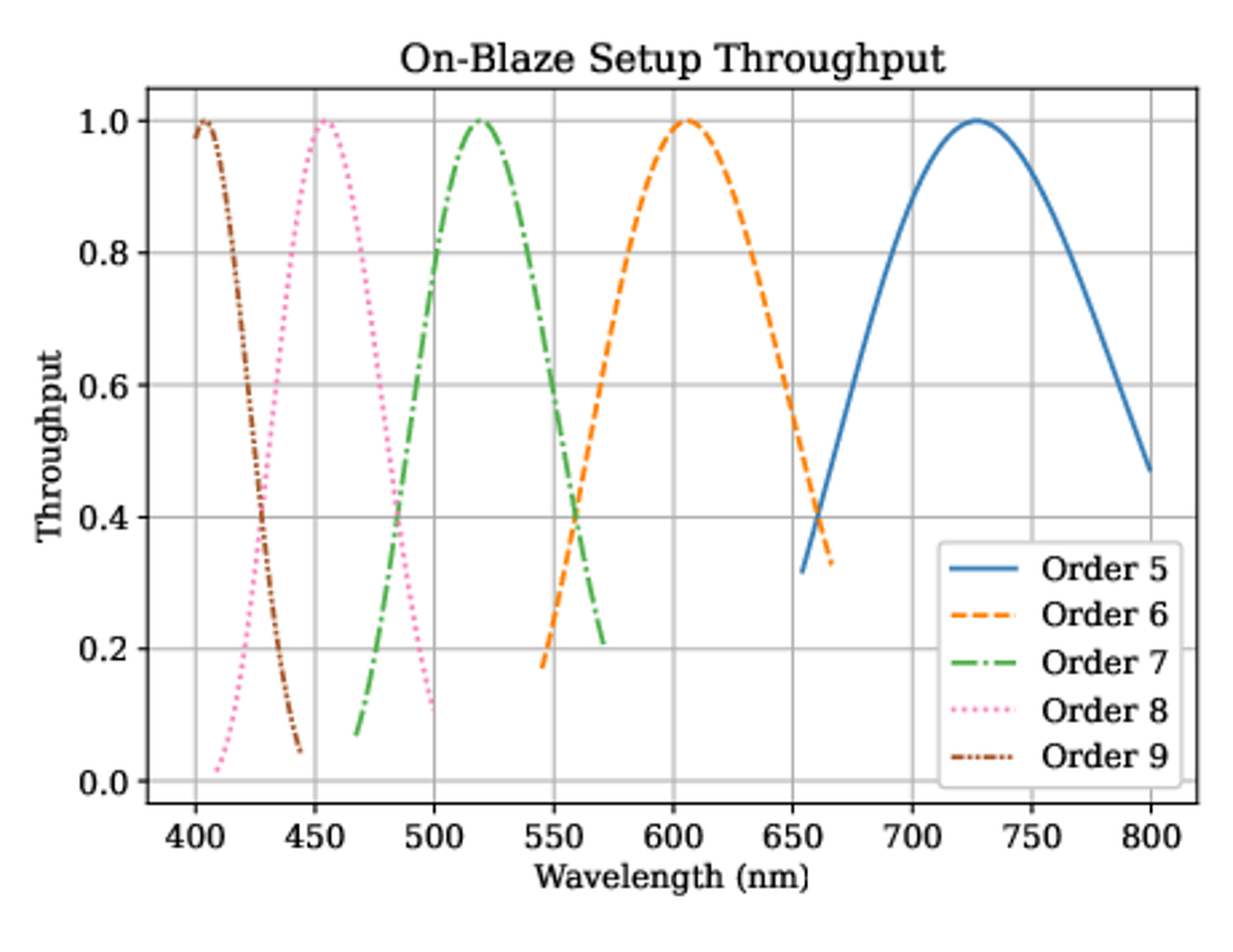}}
    \subfloat{
    \includegraphics[width=0.5\linewidth]{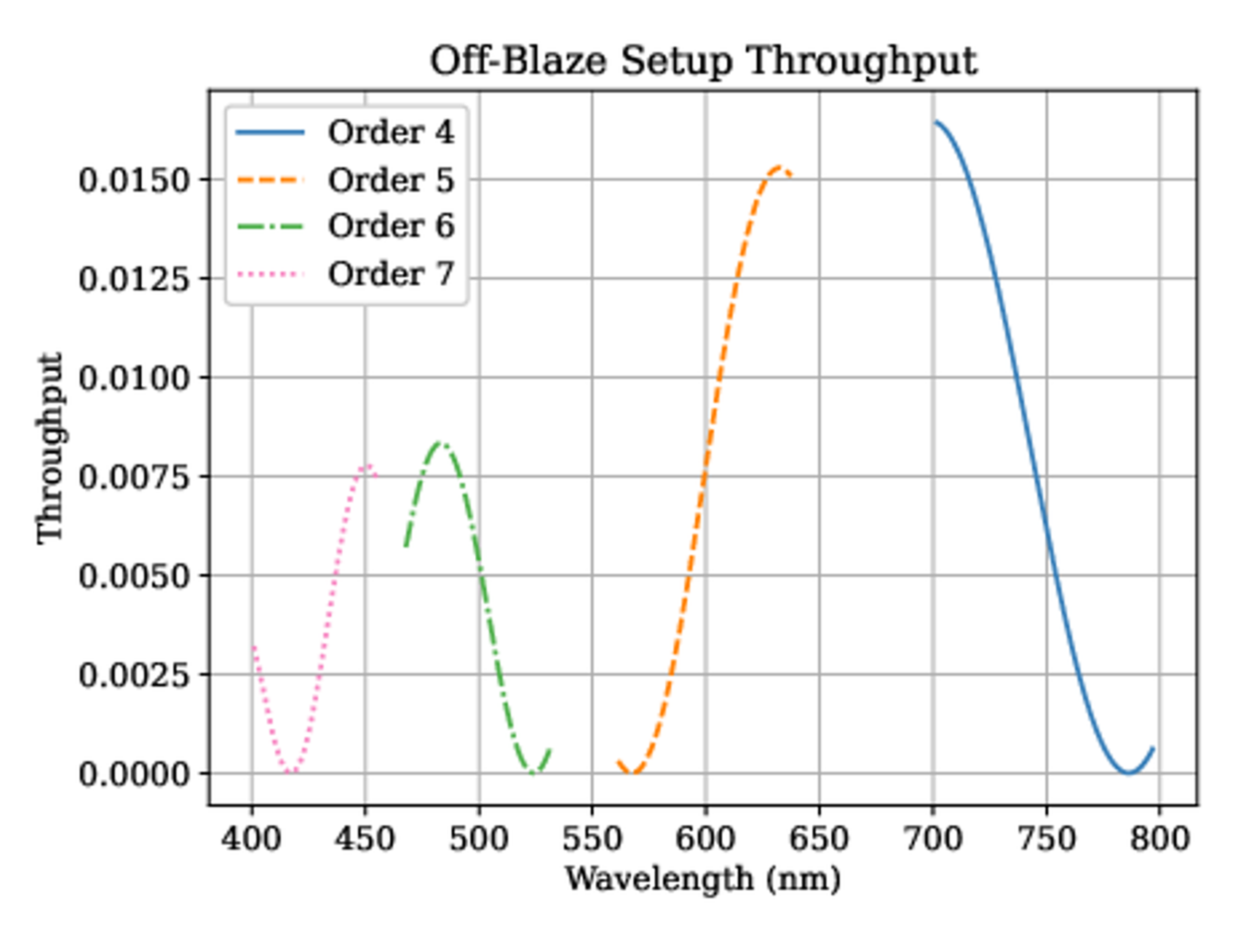}}
\caption{Grating blaze as a function of wavelength and order for (left) an on-blaze grating setup and (right) a non-ideal ``off-blaze" setup with unity peak throughput. The off-blaze setup represents the current state of \instrname. Order numbers are specific to the grating for each plot.}
\label{fig:blaze}
\end{figure*}

The simulator then applies a wavelength- and order-dependent blaze function $I(\beta(\lambda, m))$ from \citet{casini_intensity_2014} to the input spectrum $S(\lambda)$:
\begin{equation} \label{eq:blazed}
    B(\lambda, m)=S(\lambda) \cdot I(\beta(\lambda, m))
\end{equation}

The spectrum has gone from 1D in wavelength to 2D with wavelength and spectral order. Two different spectrometer configurations and their effect on the throughput are shown in Figure~\ref{fig:blaze}. While an ideal design employs an on-blaze grating configuration and full coverage of the wavelength bandpass, it can require the ruling of a custom master. The first iteration of \instrname\ is significantly off-blaze like on the right side of Figure~\ref{fig:blaze} so it was important that the simulator include such support.

\subsection{Optical Effects} \label{sec:optical}

Optical broadening is the convolution of the spectrum with the Point Spread Function (PSF) of the total instrument response. The PSF of the system is approximated here as a Gaussian Line Spread Function (LSF), $g(\lambda)$, where $\lambda_{avg}$ is the central wavelength in the instrument bandpass and $R_s$ is the design resolution. The width of the Gaussian is
\begin{equation} \label{eq:optFWHM}
    d\lambda_{\textrm{FWHM}}=\frac{\lambda_{avg}}{R_s d\lambda_{avg}}
\end{equation}
where $d\lambda_{avg}$ is the average resolution element size at $\lambda_{avg}$; this ignores a variation of about $\pm3$\% with wavelength over each order.

\begin{equation} \label{eq:optical}
    g(\lambda)=
    \frac{4\sqrt{\ln 2}}{d\lambda_{\textrm{FWHM}}\sqrt{\pi}}\exp{\left(
             -\frac{2 \text{ln}(2) \lambda^2}
             {d\lambda_{\textrm{FWHM}}^2}\right)}
\end{equation}

$B(\lambda, m)$ (Eq.~\ref{eq:blazed}) is convolved with Eq.~\ref{eq:optical} to return the optically broadened spectrum $F(\lambda, m)$.
\begin{equation}
\label{eq:optconvol}
    F(\lambda, m)
    =\int B(\tau, m)g(\tau-\lambda) d\tau
\end{equation}

\subsection{MKID Resolution} \label{subsec:MKID_R}
\begin{figure*}
    \centering
    \includegraphics[width=\linewidth]{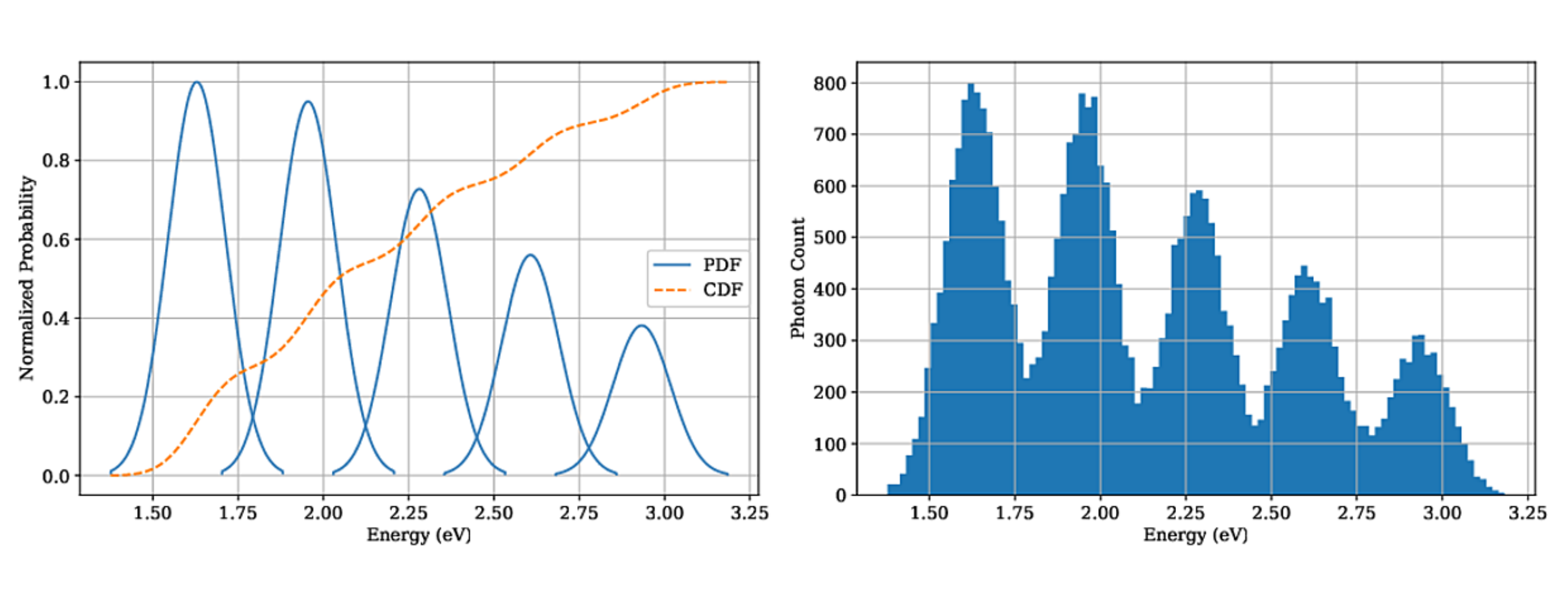}
    \caption{(left) Flux density distributions and Cumulative Density Function (CDF) of an example pixel in the on-blaze grating setup. (right) A histogram of the randomly drawn wavelengths from the CDF, using a Poisson random draw to determine the total number of photons. Each of the five Gaussian-like histogram peaks represent this MKID pixel's simulated response to the corresponding spectral order. The orders are in ascending order with energy (descending with wavelength). The simulated $\mathcal{R}_{\textrm{MKID}}=15$ is relatively low, so the orders overlap.}
    \label{fig:cdfhist}
\end{figure*}

Each order of the blazed and LSF-broadened spectrum is still physically overlapped. A secondary grating, the cross-disperser, would further separate the orders in a traditional spectrometer. Instead, the intrinsic MKID wavelength resolution is used in place of a cross-disperser. Since only one resolution element of each order overlaps on the MKID detector, the orders are separable when the MKID resolution width is smaller than the order separation. Each order is spread into a Gaussian (as in Figure~\ref{fig:cdfhist}) due to convolution with the approximately Gaussian MKID width. Each Gaussian yields the photon wavelengths from the mean, the photon distribution from the width, and the intensity from the integrated area. Every pixel contains such a Gaussian mixture distribution.

\begin{figure}[b]
    \centering
    \includegraphics[width=\linewidth]{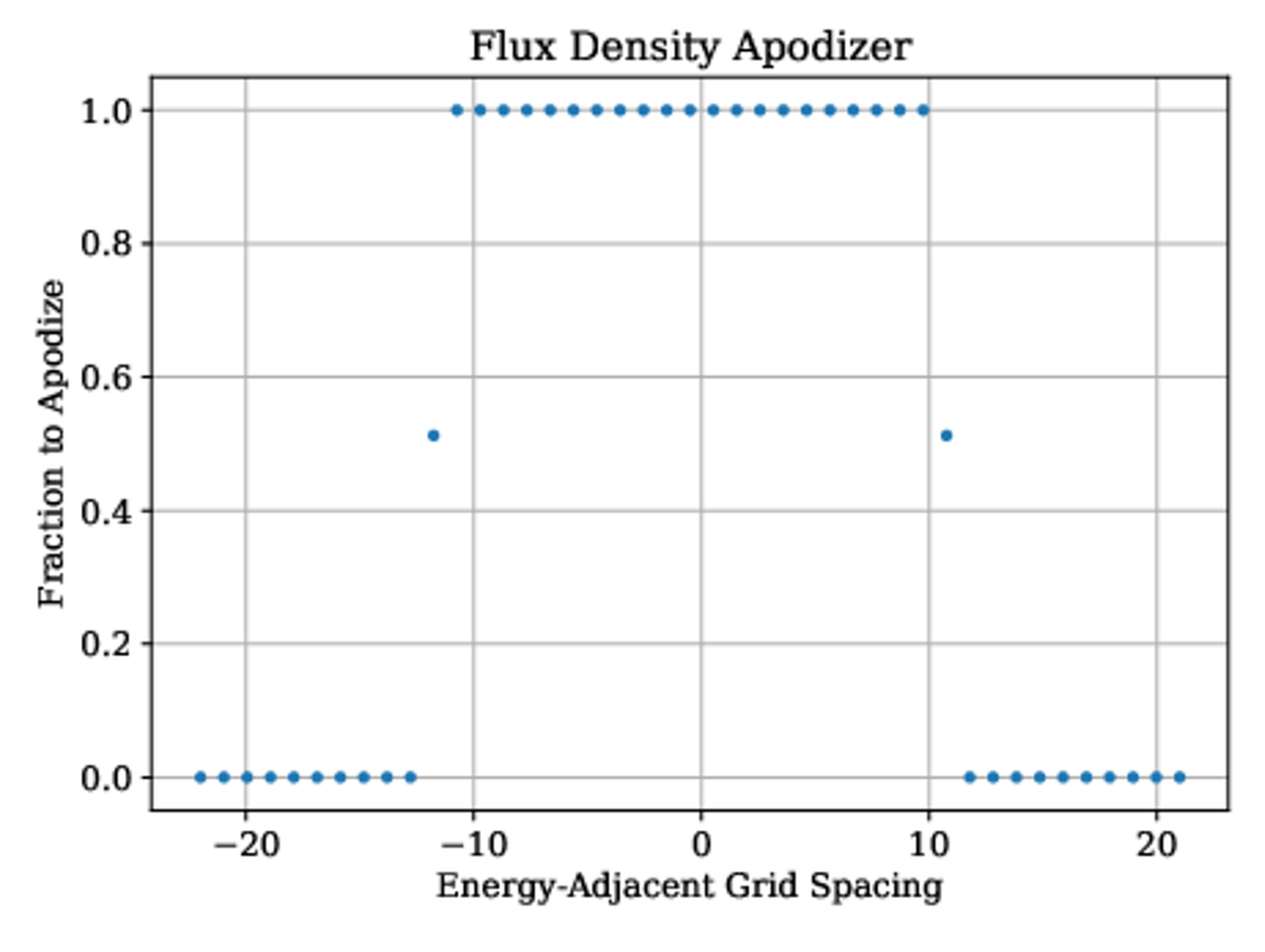}
    \caption{Multiplication with this example apodization samples the spectrum right up the edges of the pixel, where it is cut off by the fraction shown.}
    \label{fig:apod}
\end{figure}

In the simulator, the computational load of the convolution is significantly reduced by using a non-uniform grid. The grid sampling is set to ensure that even the shortest-wavelength pixel uses a well-sampled convolution kernel. Spacing for each pixel is based on that MKID pixel's resolution, where the different widths and flux densities of the pixels are handled by an apodizing mask. The flux density at each pixel is interpolated across its spectral width. Since the sampling grid has a fixed width between points, the spectral width of a pixel will only fractionally fill two of the grid points if the pixel is centered on the grid, with zero flux density filling the remaining grid points to the left and right. An example of an apodizing mask used to achieve this by multiplying it by the pixel flux density is shown in Figure~\ref{fig:apod}. The two values at about 0.2 represent the abrupt edge of the pixel in which the grid point is straddling another pixel.

The interpolated input spectrum is broadcast-multiplied with the apodizing mask. This inflates Eq.~\ref{eq:blazed} from two dimensions into three where each value corresponds to an order, pixel, and wavelength on the oversampled, apodized grid.

Next, $M_{im}(\nu)$ is the Gaussian approximation that represents the theoretical response of each order in an MKID pixel in accordance with its design resolution. It is built by mapping each pixel's central diffraction angle $\beta_i$ to the corresponding central energy $\nu_{im}$ for each order. The pixel-order standard deviation $\sigma_{im}$ is derived from the MKID FWHM $d\nu_{\textrm{MKID},im}$, defined below. $r_{pix}$ is the physical pixel size, $n_{pix}$ the number of pixels, $f$ the focal length, and $\nu_0$ is the energy for which $\mathcal{R}_{\textrm{MKID},0}$ is defined.

Since pixel-to-pixel resolution is not necessarily constant, the simulator randomizes an $\mathcal{R}_{\textrm{MKID},0,i}$ within $\pm 15\%$ of the $\mathcal{R}_{\textrm{MKID},0}$, which was loosely informed by $\mathcal{R}_{\textrm{MKID}}$ measurement variations by \citet{walter_mkid_2020}, \citet{meeker_darkness_2018}, and \citet{mazin_arcons_2013}. All three MKID arrays show an $\mathcal{R}_{\textrm{MKID}}$ variation of about $\pm2$ regardless of the wavelength measured, leading to the decision to implement $\pm 15\%$ for an instrument which is expected to have $\mathcal{R}_{\textrm{MKID}}\ge 15$. This will be updated once a working linear MKID array is characterized.

\begin{equation} \label{eq:betapix}
    \beta(i)=\alpha+\arctan\left(\frac{r_{pix}(i-n_{pix}/2)}{f}\right)
\end{equation}
\begin{equation}
\label{eq:pixel_wave}
    \nu_{im}=
        \frac{hcm}{d\left(\sin{\left(\beta(i)\right)}+\sin{\alpha}\right)}
\end{equation}
\begin{equation}
\label{eq:mkid_r}
    d\nu_{\textrm{MKID},im}=\frac{\lambda_{im}^2\nu_{im}^2}{hcR_{0,i}\lambda_0}
\end{equation}
\begin{equation}
\label{eq:M}
    M_{im}(\nu)=\frac{1}{\sigma_{im}\sqrt{2\pi}}
                 \exp{\left(-\frac{(\nu-\nu_{im})^2}{2\sigma_{im}^2}\right)}
\end{equation}
Eq.~\ref{eq:optconvol} is convolved with Eq.~\ref{eq:M} to produce a spectral flux function $D_{im}(\nu)$ which represents the theoretical energy response of each MKID pixel for each order. The limits of integration are the starting and ending energies of the resolution element, $\lambda_{im}$ and $\lambda_{i+1,m}$.
\begin{equation} \label{eq:mkidconvol}
    D_{im}(\nu)=\int_{\nu_{im}}^{\nu_{i+1,m}}F(\tau,m)M_{im}(\tau-\nu)d\tau
\end{equation}

After the convolution, the true spacing for each pixel and order pair is multiplied through the flux density spectrum in the simulator to return spectral photon flux for every order, pixel, and energy along a uniform grid, undoing the prior resampling.

In order to simulate individual photons, the cumulative density function (CDF) $H_i(\nu)$ of a pixel is computed, removing order distinction. $m_0$ is the initial order and $m_{max}$ is the final order.
\begin{equation} \label{eq:cdf}
    H_i(\nu)=
            \int_{-\infty}^\nu 
                \left[\sum_{m=m_0}^{m_{max}} D_{im}(\nu')\right]d\nu'
\end{equation}

A Poisson draw is performed on the total photon flux in each pixel to return a total number of incident photons. Each of these photons are randomly assigned an energy drawn from the CDF (Eq.~\ref{eq:cdf}) as well as a uniformly-random arrival time within the exposure time window. At this stage, the spectrum is no longer in flux, but exists as a quantized list of individual photons, each with a wavelength, timestamp, and pixel ID.

\subsection{Phase Response} \label{subsec:phase}
The MKID pixel does not directly measure energy, but a change in resonant phase $\varphi$ of its resonator. Since this $\varphi$ is approximately linear in energy $\nu$ \citep{szypryt_large-format_2017}, this simple relationship is used in the simulator:
\begin{equation}
    \varphi(\nu)=a\nu+b
\end{equation}
Each energy in the photon list is mapped to a phase response for constants $a$ and $b$.

Furthermore, the MKID phase response is not always the same for a given energy across the entire MKID array and is highly dependent on the lithography of each MKID. The same photon energy may register as $-0.5\pi$ in one MKID pixel and  $-0.55\pi$ in another. To simulate this, a random offset varying within $\pm20\%$ is multiplied through each pixel's phases, where every pixel has a different random offset. All photons in a single pixel are shifted in one direction together, not individually. This variable offset is not known without calibration. This highlights that phase is not a single, universal response to photon energy. In general, photons in the UVOIR are expected to fall between $-\pi$ and 0, where the full phase response can be from $-\pi$ to $\pi$. This is again owed to lithography; each MKID resonator is designed to be efficient for detecting specific energies.

The final product of the \instrname\ simulator is an HDF5 file containing the photon table \citep{steiger_mkid_2022} where each photon has an associated phase, timestamp, and pixel ID. This HDF5 file is functionally identical to what will be the output of the real \instrname\ instrument and has been explicitly designed this way in order to test the data reduction pipeline with known spectra.

\section{\instrnamecaps\ Data Reduction Package} \label{sec:extract}
A full \instrname\ data reduction schematic with all steps and options is shown in Appendix~\ref{sec:package}. The alphabetical labels are referenced here for the relevant steps. The \instrname\ data reduction takes three MKID photon tables: one each for order-sorting, wavelength calibration, and target observation. In the MSF-retrieval step (A) a continuum source photon table (X) is binned and fit with $n_{orders}$ Gaussians at each pixel; virtual pixel bins are determined; covariance between orders calculated (B); and the MKID Spread Function (MSF) is saved to file (C). In the wavelength calibration step (D) an emission line photon table (Y) is order-sorted and bleed-corrected using the MSF (E); it is saved to a FITS file (F), wavelength calibrated (G); and the dispersion solution saved to file (H). Finally, in the extraction step (J) a target observation photon table (Z) is order-sorted and bleed-corrected using the MSF (E); it is saved to a FITS file (K), the dispersion relation is applied (L); and the final, extracted spectrum is saved to the previous FITS file (M).

The data reduction is designed to be usable on photon tables generated from either the simulator or a real \instrname\ instrument.

\subsection{MKID Spread Function} \label{subsec:msf}
A sufficiently high SNR ($>50$) continuum source is required in order to recover the MKID Spread Function (MSF) because all orders must contain enough flux and not be dominated by noise to distinguish one from the other. Since the MSF relies only on a continuum source, such as a tungsten lamp, SNR may be increased via longer integration if needed.

A file that already contains the order-sorting MSF calibration may be supplied instead if a relevant calibration has already been conducted. This means that not every science observation is required to undergo its own MSF calibration, which can be time-consuming both during the night and in post-processing.

\subsubsection{The Histogram} \label{subsub:hist}
\begin{figure}
\subfloat{
    \includegraphics[width=\linewidth]{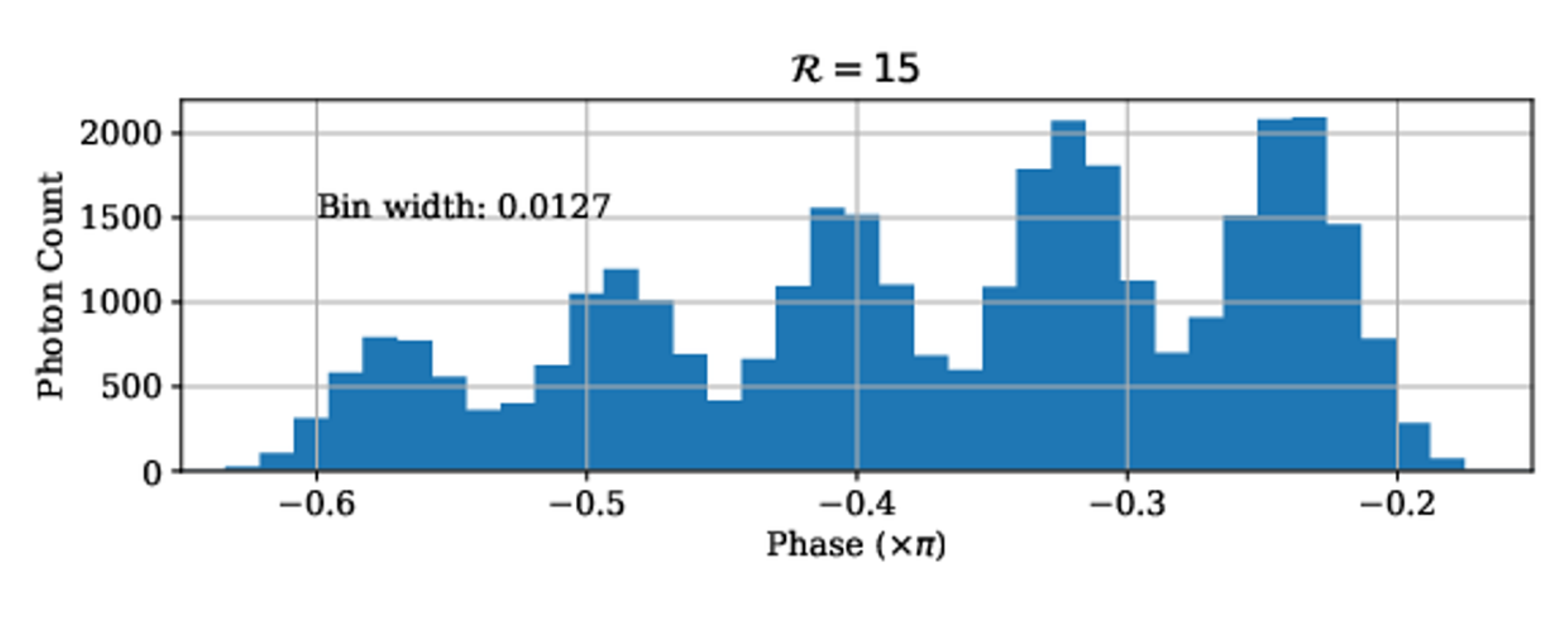}}\\
    \subfloat{
    \includegraphics[width=\linewidth]{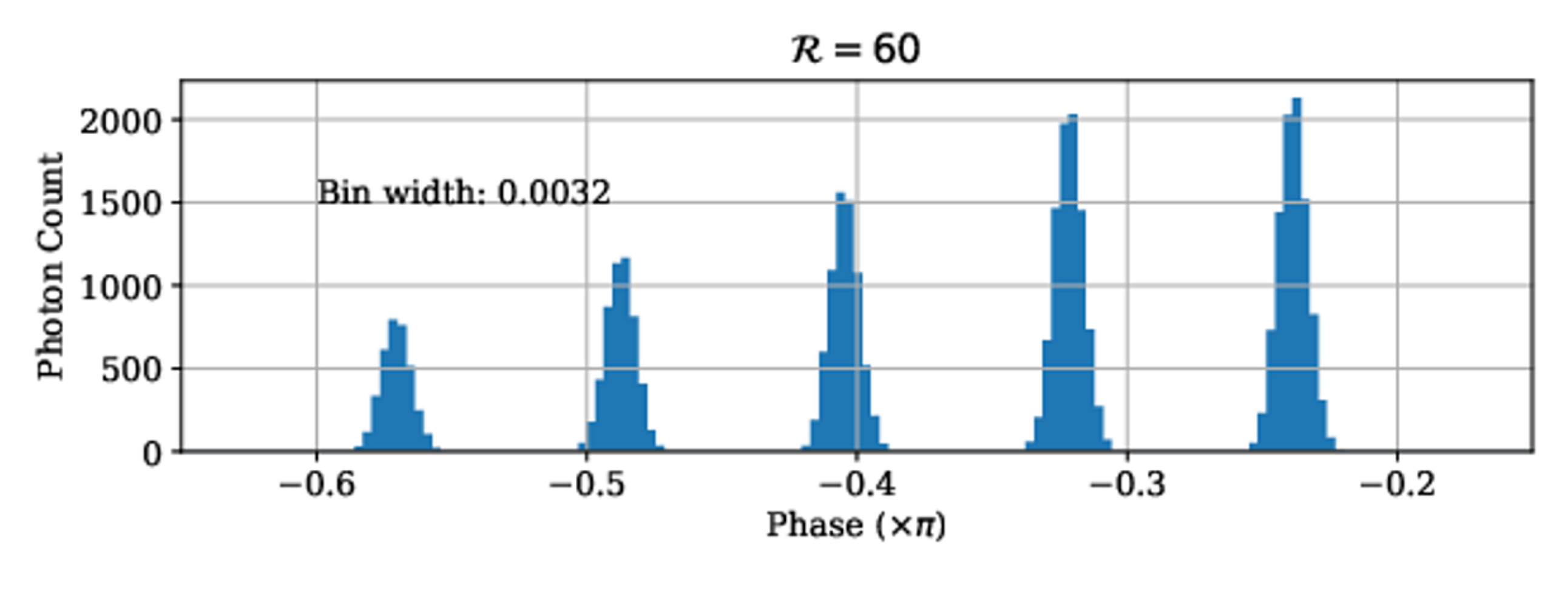}}
    \caption{Difference in bin width and distribution of photons between similar \instrname\ simulations with $\mathcal{R}_\textrm{MKID}$ of 15 (top) and 60 (bottom), for the same pixel.}
    \label{fig:binwidth}
\end{figure}

Since the photon table is a quantized list of individual photons, they must be binned before the MSF can be fit. Each order's expected MKID phase response distribution in any one pixel is approximately Gaussian. There must be enough bins across the distribution to clearly identify this shape. For example, with ten bins across $6\sigma$, the slope up and down the Gaussian peak can be distinguished. More bins would necessitate longer integration to reduce noise. With too few, fine details would be lost. The bin width was calculated to be a function of $\mathcal{R}_\textrm{MKID}$. Because the MKID phase response is approximately linear with energy, $\mathcal{R}_\textrm{MKID}$ can be translated to a bin width in phase via $\mathcal{R}_\textrm{MKID}=\nu/d\nu\approx\varphi/d\varphi$. Using ten bins across $6\sigma$ of the average $\mathcal{R}_\textrm{MKID}$ still allows lower resolution pixels to be sampled with 7-8 bins. Figure~\ref{fig:binwidth} shows this effect for two different $\mathcal{R}_\textrm{MKID}$.

\subsubsection{The Fitting Function} \label{subsub:fitfunc}
With the data binned, the separation between orders becomes clear to the naked eye. But to separate orders in thousands of pixels automatically, a physically-motivated, well-constrained function is defined to fit to this data based on the Gaussian approximation (Eq.~\ref{eq:M}) from the simulator.

An unknown initial parameter $\varphi_{im_0}$ is defined to represent the pixel $i$ peak location of the 1st order, $m_0$. Then, a pixel-specific second-order polynomial of photon energy $\nu_i(\varphi)$ is define as a function of phase $\varphi$ with unknown coefficients $e_n$.
\begin{equation}
    \nu_i(\varphi)=e_0+e_1\varphi+e_2\varphi^2
\end{equation}

The next step is to constrain $\nu_i$ and the remaining $\varphi_{im}$ with the fundamental grating equation:

\begin{equation}
   \nu_i(\varphi_{im'})=\frac{m'}{m_0} \nu_i(\varphi_{im_0})\\
\end{equation}

At the $m_0$ order phase $\varphi_{im_0}$, there is some associated, unknown energy $\nu_i(\varphi_{im_0})=\nu_{m_0}$. This is divided out and absorbed into the coefficients to eliminate a degenerate fitting parameter.

Physically-motivated constraints must now be applied to the energy coefficients. Since $\varphi_{im_0}$ is the peak with the smallest phase value (lowest energy corresponding to the lowest order), this free parameter is constrained to $\pm0.2\pi$ of it's initial guess value. $e_1$ is the slope condition of the phase-energy relationship. MKIDs return more negative phases for higher energy photons, so $e_1$ is constrained to only negative values. $e_2$ is the quadratic term and is arbitrarily constrained to the $\pm 10^{-2}$ regime since the energy-phase relationship is approximately linear.

Next, another second-order polynomial $S_i(\nu)$ is defined with unknown coefficients $s_n$ as the standard deviation of the pixel response as a function of energy:
\begin{equation}
    S_{i}(\nu_i(\varphi))=s_0+s_1\nu_i(\varphi)+s_2\nu_i(\varphi)^2
\end{equation}

The standard deviation parameters can be constrained similarly to the energy parameters. $s_0$ is the y-intercept of the energy-standard deviation relationship. Since standard deviation must be positive, $s_0$ is constrained to positive values. From Eq.~\ref{eq:mkid_r}, lower energy photons have lower R, which corresponds to larger standard deviation, so $s_1$ is constrained to negative values. Like the energy polynomial, this relationship is approximately linear, so $s_2$ is also constrained to $\pm 10^{-2}$.

$\varphi_{im}$, $S_{i}(\nu_i(\varphi_{im}))$, and the independent, unknown amplitude parameters $A_{im}$ represent the Gaussian mean, standard deviation, and amplitude for the discrete integer orders $m$ and integer pixel indices $i$.

Finally, the objective function is:
\begin{equation} \label{eq:fitfuncorders}
    G_{im}(\varphi)=A_{im}\exp{\left[-\frac{1}{2}\left(\frac{\varphi-\varphi_{im}}{ S_i(\nu_i(\varphi_{im}))}\right)^2\right]}
\end{equation}
\begin{equation} \label{eq:fitfunc}
    G_{i}(\varphi)=\sum_{m=m_0}^{m_{max}}G_{im}(\varphi)
\end{equation}

The metric to be minimized is the weighted, reduced $\chi_i^2$ for each pixel. The extensive functionality of \texttt{lmfit} \citep{newville_lmfit_2016} is used to conduct non-linear least squares fitting with the aforementioned parameter constraints. This fit is conducted individually for each pixel. Initial guesses of the Gaussian mean and width are obtained from the quantized data using a clustering algorithm, then the corresponding amplitude guess at the mean is derived from the binned data. Once fitting has concluded, full expressions for the energy $\nu_i(\varphi_i)$ and MKID standard deviation $S_i(\varphi_i)$ will be returned.

For a continuum source with an average SNR of at least 50, the typical MSF least squares fitting uncertainty on non-zero parameters is 0.7\%. These parameter uncertainties translate to an uncertainty on the magnitude of the adjacent order-bleed (noise) that is on average less than 0.01\% at $\mathcal{R}_{\textrm{MKID}}=15$.  A continuum source may be observed for as long as needed to achieve the necessary MSF uncertainty and SNR.

\subsubsection{Virtual Pixels} \label{subsub:virtualpixels}
\begin{figure}
    \centering
    \includegraphics[width=\linewidth]{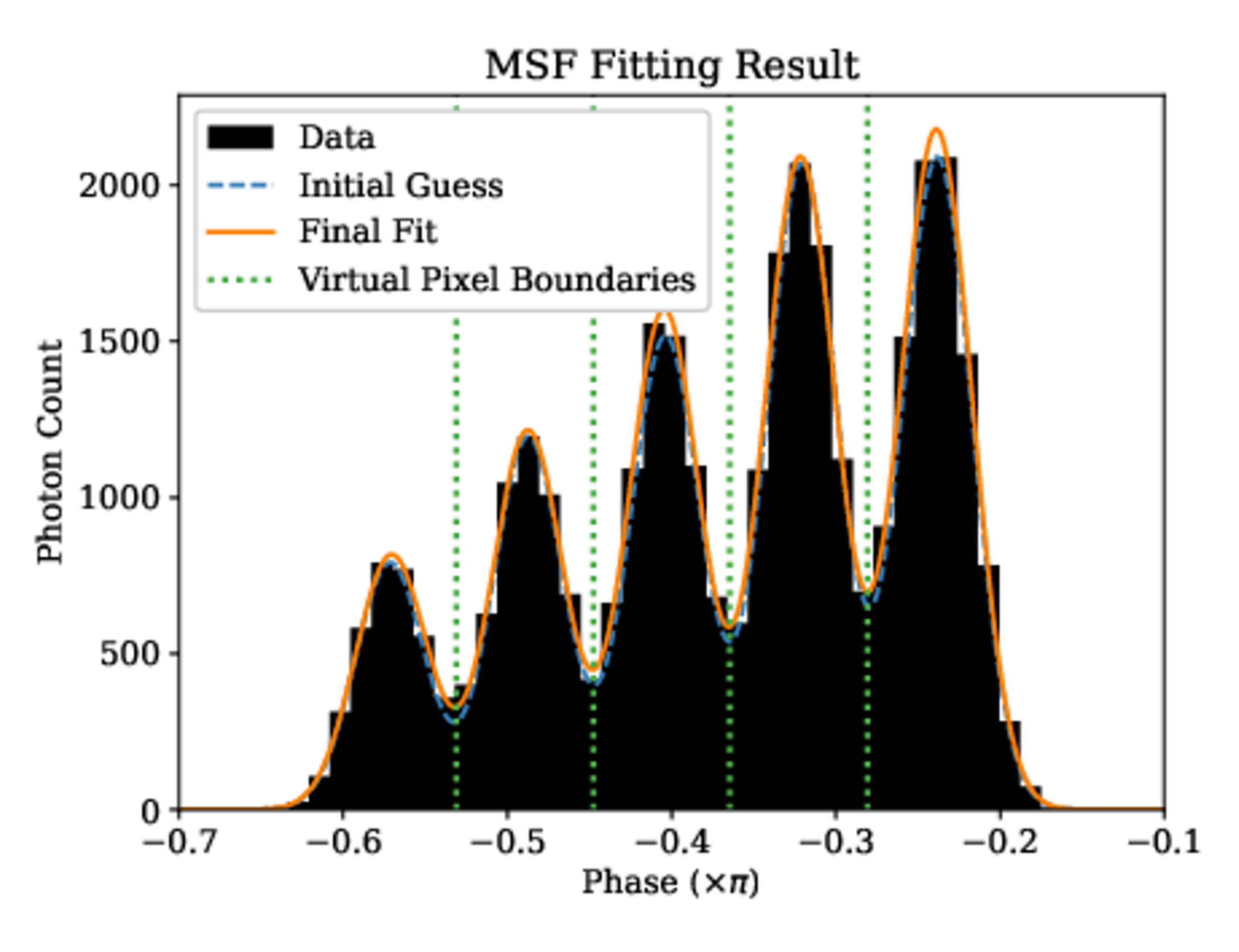}
    \caption{Example phase histogram in the on-blaze grating setup for $\mathcal{R}_{\textrm{MKID}}=15$ with the initial guess, converged fitting function $G_i$, and virtual pixel (order) boundaries.}
    \label{fig:edges}
\end{figure}

The purpose of fitting $n_{order}$ Gaussians is to subdivide a single MKID pixel into multiple virtual pixels, each of which represent one spectral order. The points of intersection of the $G_{im}$ and $G_{im'}$ functions become the virtual pixel boundaries. Any counts "lost" to the left of the boundary by the Gaussian on the right is part of its approximate order-bleed fraction, and vice versa. An example pixel with boundaries is shown in Figure~\ref{fig:edges}.

Integrating the counts within each boundary for all pixels returns all $n_{orders}\times n_{pixels}$ counts $C_{\textrm{bound}}$. 

\subsubsection{Covariance} \label{subsub:covar}
Order bleeding is the phenomenon whereby the virtual pixel boundary effectively slices away some of the count in each order and groups it into an adjacent order. If one order were significant brighter than an adjacent one, and the MKID resolution was poor or the peak separation was small, a large number of photons from the brighter order would be counted as part of the dimmer one. Consequently, false "emission" lines from very bright lines would populate the spectrum in the adjacent order that do not actually exist. To take this into consideration, a covariance matrix of the orders is calculated, where covariance here refers to the estimated fraction of each order that has been counted another. The fraction of order $m_k$ being counted as part of order $m_\ell$ is recorded in each $k$-$\ell$ position. Using Eq.~\ref{eq:fitfuncorders}, each element of the covariance matrix is:
\begin{equation}
    \kappa_{m_km_\ell}=
        \frac{\sum_{\varphi=\phi_{m_\ell-1}}^{\phi_{m_\ell}}
                G_{m_k}(\varphi)}
             {\sum_{\varphi=\varphi_{min}}^{\varphi_{max}}
                G_{m_k}(\varphi)}
\end{equation}
where $\phi_{m_\ell-1}$ is the left virtual pixel boundary, $\phi_{m_\ell}$ is the right virtual pixel boundary, and $\varphi_{min}$ and $\varphi_{max}$ are the boundaries of the entire phase space.

The matrix is:
\begin{equation}
    K_i=
    \begin{bmatrix}
        \kappa_{i,m_0m_0}&\kappa_{i,m_0m_1}&\hdots&\kappa_{i,m_0m_{max}}\\
        \kappa_{i,m_1m_0}&\kappa_{i,m_1m_1}&\hdots&\kappa_{i,m_1m_{max}}\\
        \vdots         & \vdots        & \ddots & \vdots       \\
        \kappa_{i,m_{max}m_0}&\kappa_{i,m_{max}m_1}&...&
        \kappa_{i,m_{max}m_{max}}
    \end{bmatrix}
\end{equation}

\subsection{Order Count Correction} \label{subsec:bleed}
The virtual pixel boundaries and covariance matrix must then be applied to the emission lamp and observed target photon tables. The virtual pixel boundaries given by the MSF bins each photon table into a two-dimensional spectrum of counts.

\begin{figure*}
    \centering
    \subfloat{
    \includegraphics[width=0.33\linewidth]{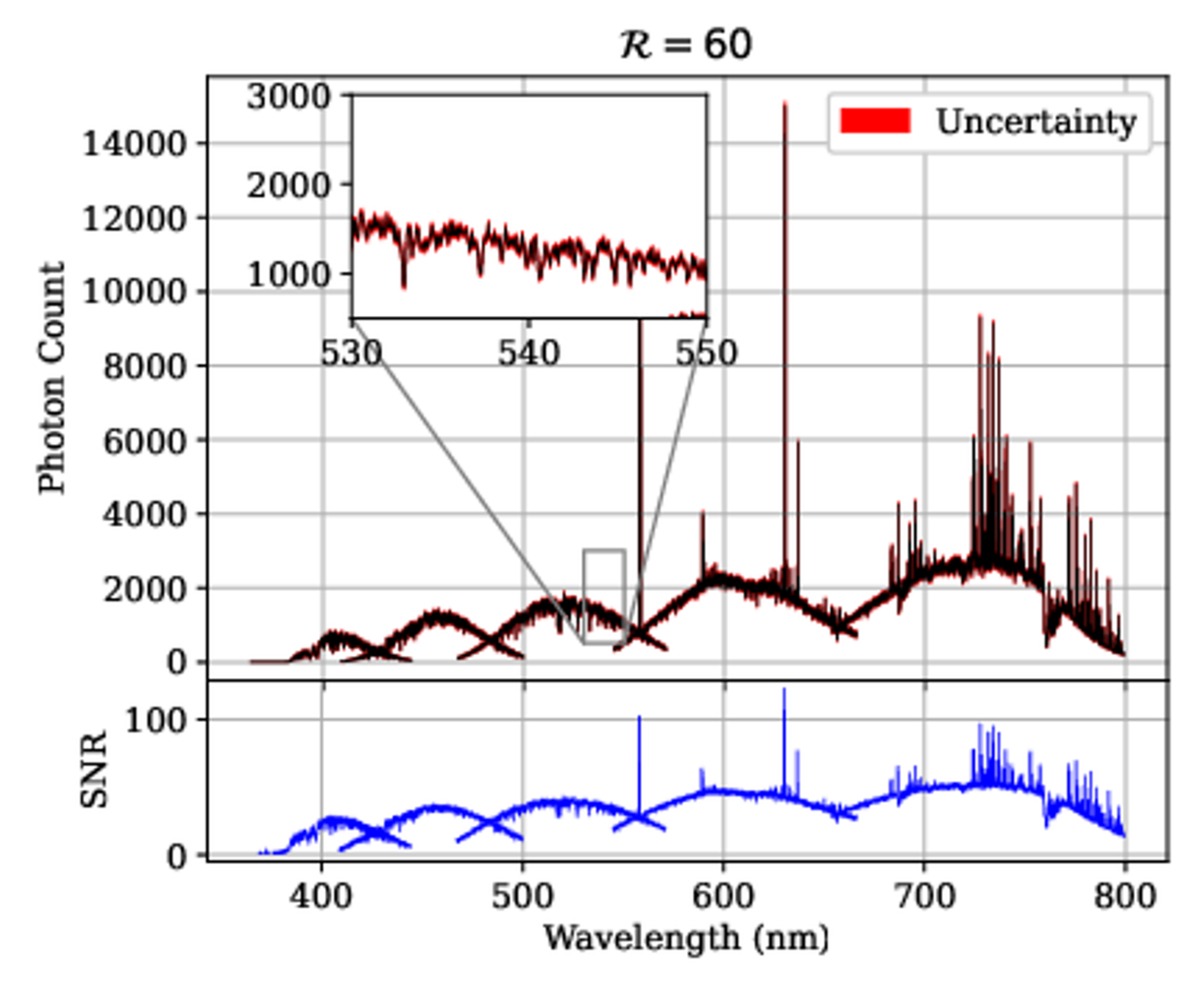}}
    \subfloat{
    \includegraphics[width=0.33\linewidth]{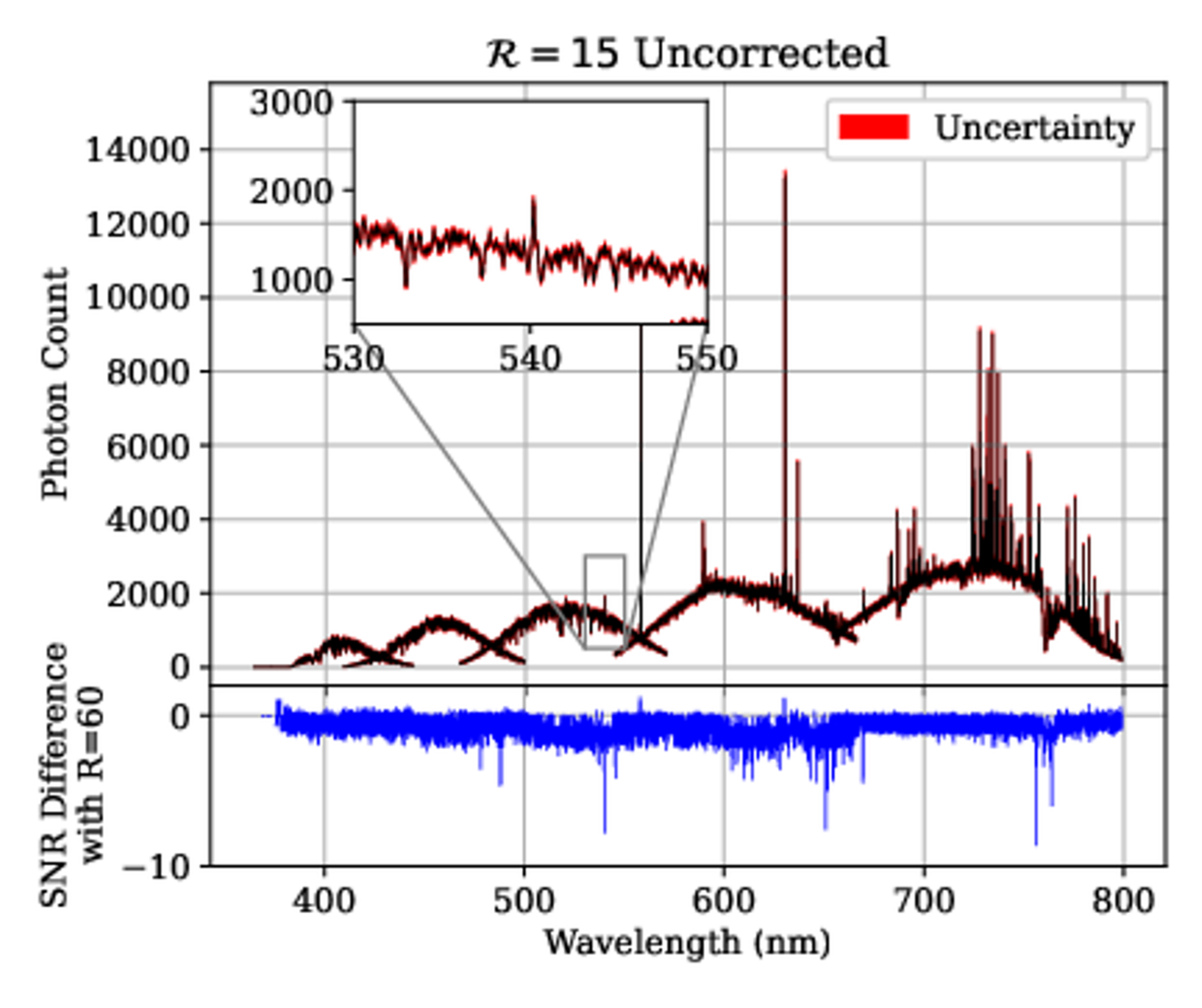}}
    \subfloat{
    \includegraphics[width=0.33\linewidth]{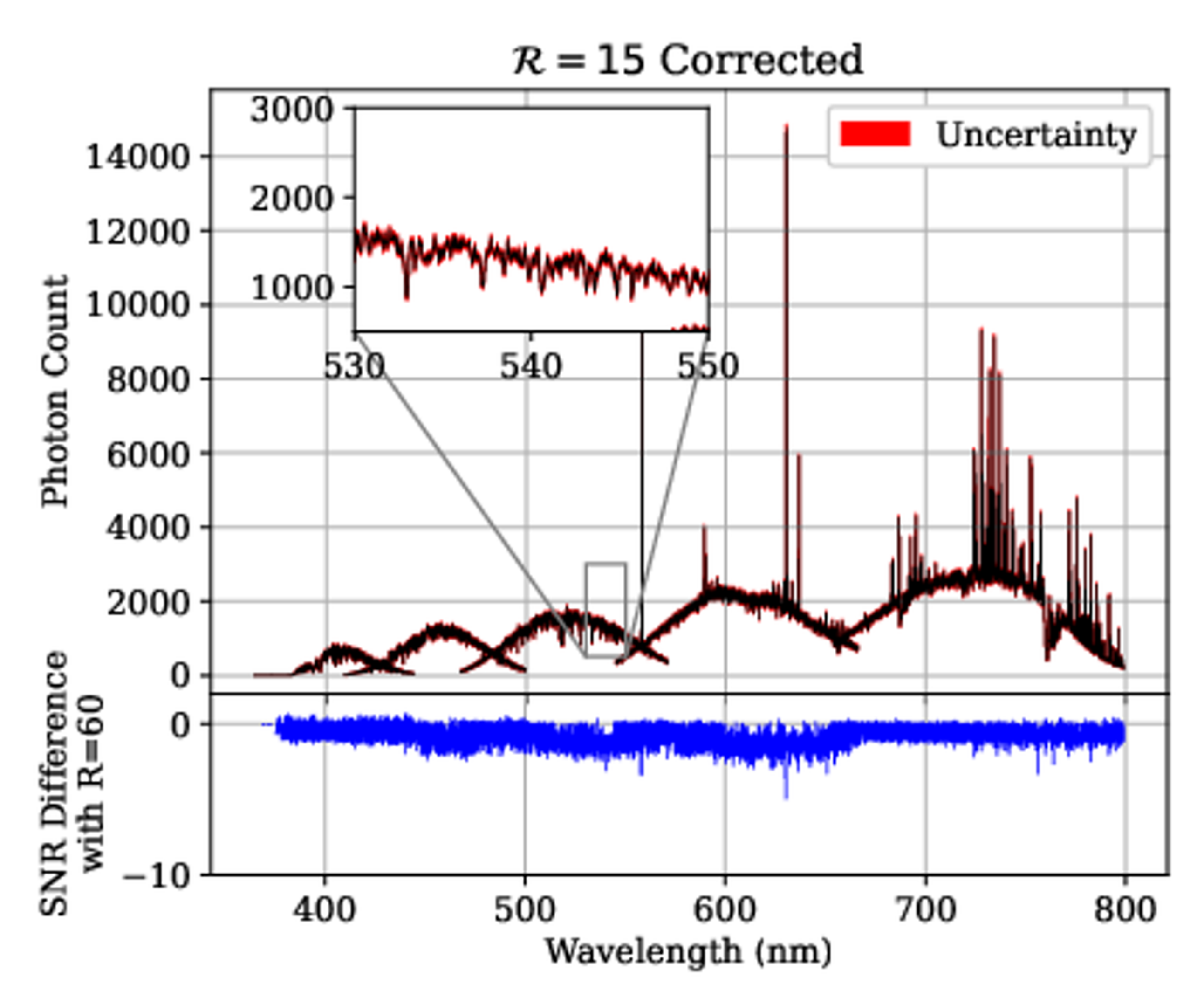}}
    \caption{Effect of bleeding subtraction using a PHOENIX G5 spectrum with prominent airglow. (left) On-blaze configuration with $\mathcal{R}_{\textrm{MKID}}=60$ and virtually no overlap. The strong emission lines do not bleed into other orders and no MSF correction was done. (middle) The same exact configuration except $\mathcal{R}_{\textrm{MKID}}=15$, also with no correction. The very bright line near 630~nm has bled into the adjacent order shown in the inset. The residual plot shows the difference in SNR from $R=60$. As a result, the SNR is lower than average where there is significant bleed. (right) Order bleeding correction from the MSF after being applied to the middle plot. The bleed has been reduced and the $\pm$ correction has been propagated to the $\mp$ uncertainty. The bleed-heavy, low SNR points have been smoothed out with lowered airglow noise. Both simulations used the same random seed.}
    \label{fig:orderbleed}
\end{figure*}

Each spectrum must now be individually corrected for order-bleeding from the MSF covariance and have this correction be propagated through to the uncertainty. The corrected counts for each order and pixel $C_{\textrm{true}}$ are retrieved from the bounded counts $C_{\textrm{bound}}$ with

\begin{equation}
    \begin{bmatrix}
        C_{\textrm{true}, im_0}\\
        C_{\textrm{true}, im_1}\\
        \vdots\\
        C_{\textrm{true}, im_{\textrm{max}}}\\
    \end{bmatrix}
    =
    \begin{bmatrix}
        C_{\textrm{bound}, im_0}\\
        C_{\textrm{bound}, im_1}\\
        \vdots\\
        C_{\textrm{bound}, im_{\textrm{max}}}\\
    \end{bmatrix}
    K^{-1}_i
\end{equation}

An example of this correction is shown for strong night sky emission lines in a dim PHOENIX spectrum (Figure~\ref{fig:orderbleed}). The bleed value $\mu$ was then combined with Poisson noise $N$ to return the uncertainty on each order and pixel:
\begin{equation}
    \sigma = \sqrt{\mu_{\kappa}+N_{i,m}}
\end{equation}

As seen in Figure~\ref{fig:edges}, $\mathcal{R}_{\textrm{MKID}}=15$ results in an extraction with slightly overlapping solution functions. With $\mathcal{R}_{\textrm{MKID}}<15$, bleeding across orders will increase. In a continuum source simulation with $\mathcal{R}_{\textrm{MKID}}=12$, the average adjacent order bleed was 7.2\%. For comparison, $\mathcal{R}_{\textrm{MKID}}=15$ yields an average bleed of 3.5\% and $\mathcal{R}_{\textrm{MKID}}=60$ gives $10^{-9}$\%. To keep bleed below 5\%, $\mathcal{R}_{\textrm{MKID}}=15$ is the rough lower limit for this configuration with five spectral orders, though the eventual science goals will guide this. The minimum $\mathcal{R}_{\textrm{MKID}}$ would be much higher for a configuration packing, say, 20 orders into the same phase space.

\subsection{Wavelength Calibration} \label{subsec:wavecal}
In the wavelength calibration step, a photon table with emission lamp data is needed to match pixels to wavelengths, though a file that already contains the solution may be supplied instead.

PyReduce \citep{piskunov_optimal_2021} is repurposed to complete the wavelength calibration. Several line atlases from the NIST Database \citep{kramida_atomic_2009} have already been retrieved and are available in the package. Other atlases can be downloaded and called in the data reduction script.

\begin{figure}
    \centering
    \includegraphics[width=\linewidth]{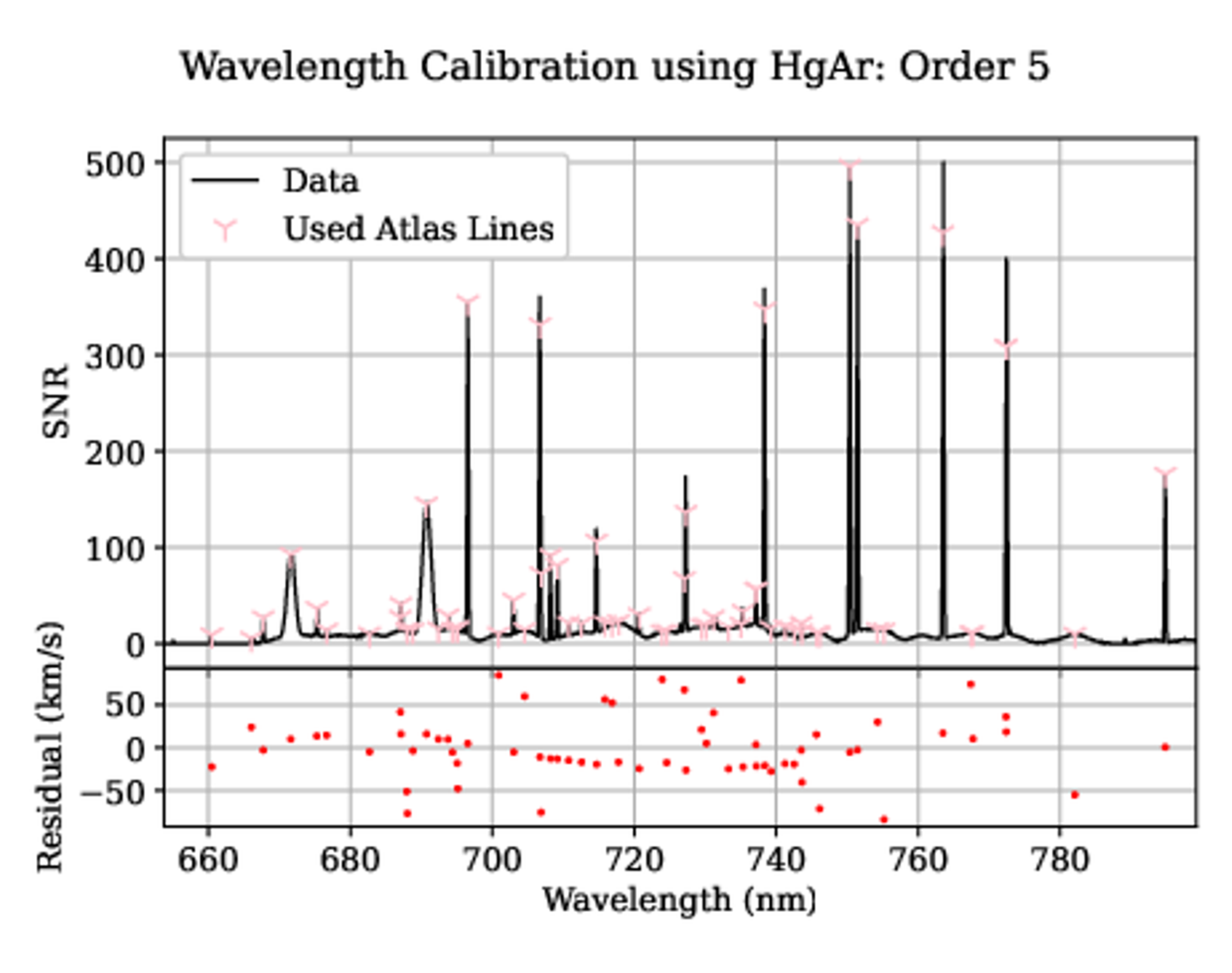}
    \caption{A PyReduce wavelength calibration comparing the used HgAr atlas lines and one order of the HgAr lamp spectrum in the on-blaze grating setup. The polynomial degree for the solution is four and the residual between the wavelengths is also shown. The residual upper limit was set to $85$ km/s, which corresponds to the width of about one MKID pixel.}
    \label{fig:wavecal}
\end{figure}

In a line-by-line fashion, PyReduce compares the atlas to the MSF-binned emission lamp spectrum along with the initial wavelength guess for each pixel that can be derived from the grating equation and spectrometer properties. The result is a fit of the virtual pixel indices to wavelength as a polynomial, where the polynomial degree is up to the user. A higher degree lets the wavelength axis squeeze and stretch to the pixel indices if the relationship is expected to be highly nonlinear. It will usually return smaller residuals and not discard as many lines as lower degrees. Lower degrees are more resistant to runaway (unconstrained) behavior in regions without good emission line data. Figure~\ref{fig:wavecal} shows an example comparison between the theoretical and calibrated wavelengths of a single order.

The binned target observation is then trivially paired with the dispersion solution and saved to a FITS file.

\section{Conclusion} \label{sec:conclusion}
The \instrname\ simulator provides a glimpse of what to expect from this novel instrument. Its mathematically- and physically-motivated design ensures that output will resemble realistic data as much as possible. It has also been instrumental in the development of the \instrname\ data reduction package. As a result, analysis on instrument output could potentially begin immediately; accelerating the pace with which pipeline improvements and physical upgrades can be made. In particular, the degree to which orders bleed into one another can be characterized via Gaussian fitting to facilitate bleeding correction; a method that transfers uncertainty and false features in the spectrum to its uncertainty.

The \instrname\ simulator and data reduction can be integrated with output from both existing and proposed telescopes as a performance showcase to make the argument that MKID-based spectrometers are a competitive alternative to traditional spectrometers.

\begin{acknowledgments}
This work is made possible by the National Science Foundation, grant number 2108651, which supported C. Kim and Dr. Bailey. Kim is also supported by a NASA Space Technology Graduate Research Opportunity, grant number 80NSSC23K1220. Dr. L\'opez is supported by the NSF Postdoctoral Fellowship, grant number 2304168.
\end{acknowledgments}

\software{numpy \citep{van_der_walt_numpy_2011},
          scipy \citep{virtanen_scipy_2020},
          astropy \citep{astropy_collaboration_astropy_2013},
          synphot \citep{stsci_development_team_synphot_2018},
          lmfit \citep{newville_lmfit_2016},
          PyReduce \citep{piskunov_optimal_2021}}

\bibliography{main}{}
\bibliographystyle{aasjournal}

\clearpage
\appendix
\section{Simulator} \label{sec:simoptions}

\begin{figure}[H]
    \centering
    \includegraphics[width=0.9\linewidth]{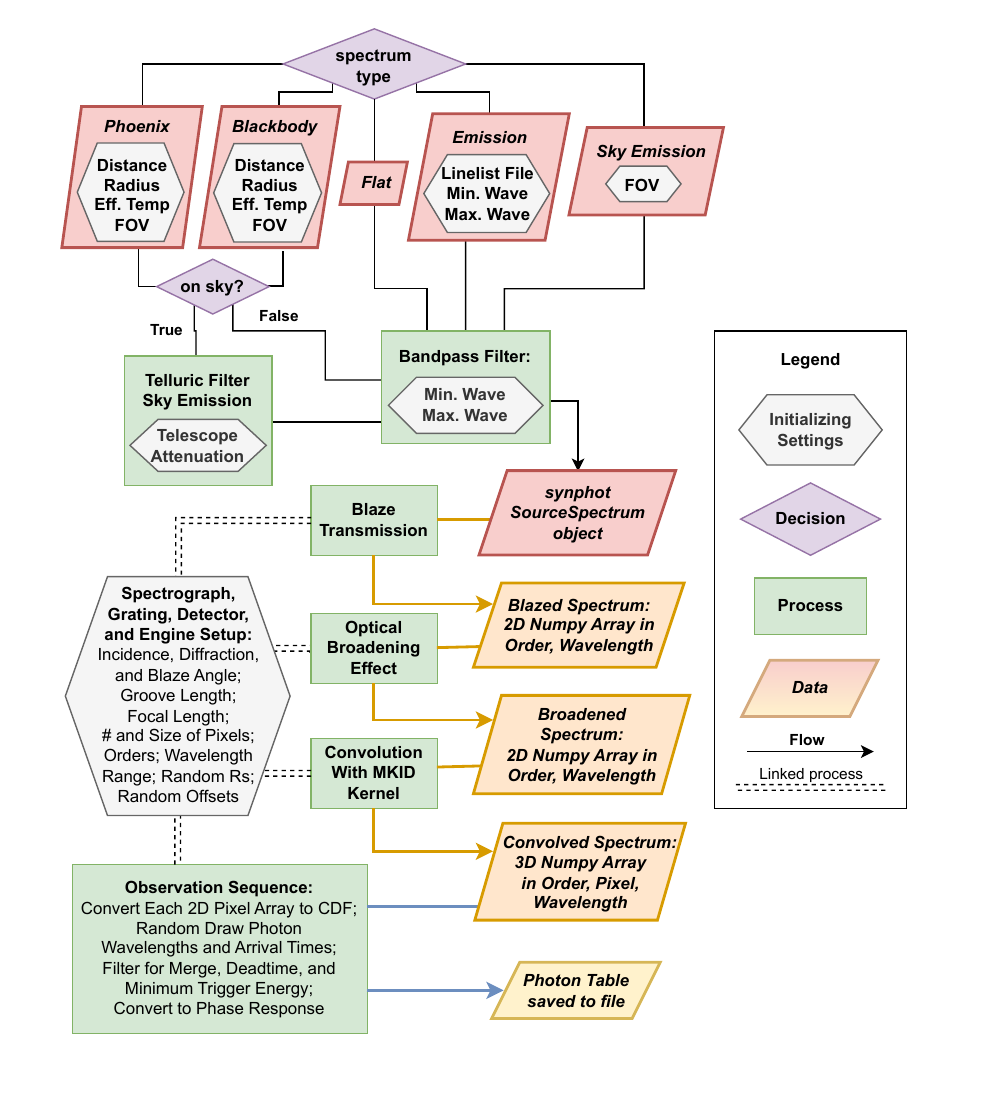}
    \caption{A flow chart schematic detailing how and where \instrname\ simulator steps and options are implemented.}
    \label{fig:simschematic}
\end{figure}

\section{Data Reduction Package} \label{sec:package}

\begin{figure}[H]
    \centering
    \includegraphics[width=\linewidth]{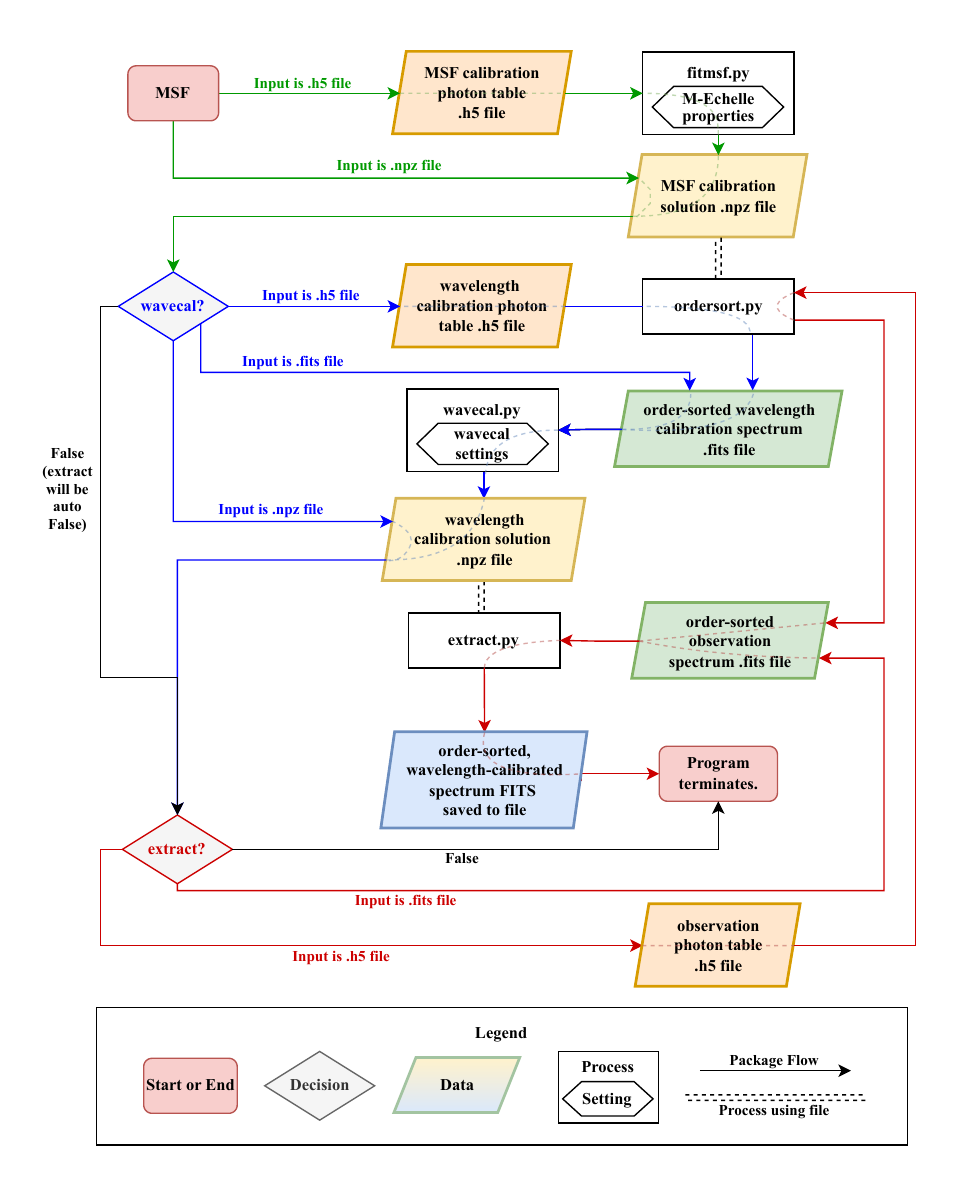}
    \caption{A flow chart schematic detailing how and where \instrname\ data reduction steps and options are implemented.}
    \label{fig:packageschematic}
\end{figure}

\end{document}